\documentclass[aps,prb,twocolumn,showpacs,amsmath,amssymb,amsthm,superscriptaddress,groupedaddress]{revtex4}
\usepackage{graphics}
\usepackage{epsfig}
\usepackage[colorlinks=true,citecolor=blue,linkcolor=red,linktocpage=true,pagebackref=false]{hyperref}
\usepackage{soul} 
\usepackage[usenames, dvipsnames]{color}
\usepackage{braket}
\usepackage[euler]{textgreek}

\begin{document}
\title{
High-temperature fusion of a multielectron leviton
}
\author{Michael Moskalets}
\email{michael.moskalets@gmail.com}
\affiliation{Department of Metal and Semiconductor Physics, NTU ``Kharkiv Polytechnic Institute", 61002 Kharkiv, Ukraine}

\date\today
\begin{abstract}
The state of electrons injected onto the surface of the Fermi sea depends on temperature. 
The state is pure at zero temperature and is mixed at finite temperature. 
In the case of a single-electron injection, such a transformation can be detected as a decrease in  shot noise with increasing temperature. 
In the case of a multi-electron injection, the situation is more subtle. 
The  mixedness helps the development of quantum-mechanical exchange correlations between injected electrons, even if such correlations are absent at zero temperature. 
These correlations enhance the shot noise, what in part counteracts  the  reduction of noise with temperature. 
Moreover, at sufficiently high temperatures, the correlation contribution to noise predominates over the contribution of individual particles.
As a result, in the system of $N$ electrons, the  apparent charge (which is revealed via the shot noise) is changed from $e$ at zero temperature to $Ne$ at high temperatures.  
It looks like the exchange correlations glue up electrons into one particle of  total charge and energy.   
This point of view is supported by both charge noise and heat noise. 
Interestingly, in the macroscopic limit, $N\to \infty$, the correlation contribution completely suppresses the effect of temperature on noise. 
\end{abstract}
\maketitle

\section{Introduction}
\label{sec1}

The recent advances in creation \cite{Blumenthal:2007ho,Feve:2007jx,Kaestner:2008gv,{Fujiwara:2008gt},{Roche:2013jw},Dubois:2013dv,{Rossi:2014kp},{Tettamanzi:2014gx},{dHollosy:2015ez},{vanZanten:2016fl}} and characterization \cite{{Leicht:2011ke},{Fletcher:2012te},Bocquillon:2013dp,Dubois:2013dv,{Jullien:2014ii},Freulon:2015jo,{Waldie:2015hy},{Vanevic:2016eq},{Ryu:2016kb},{Marguerite:2017tn}} of single-electron wave-packets prepare the basis for the engineering of multi-electron quantum states. 

The first steps in this directions were already done. 
For instance, the wave spitter was used as a passive two-particle source. \cite{Dubois:2013dv,Bocquillon:2013dp,{Bocquillon:2013fp},Marguerite:2016ur,Glattli:2016wl} 
A dynamic quantum dot \cite{Blumenthal:2007ho,{Leicht:2011ke},{Fricke:2012vk},Fletcher:2012te,{Ubbelohde:2014vx},Waldie:2015hy}  and a Lorentzian voltage pulse \cite{Dubois:2013dv,{Glattli:2016tr}}  were used as an active multi-particle source. 

Note also the theoretical proposals of multi-particle sources. 
A two-electron source can be built from the two quantum capacitors\cite{Buttiker:1993wh,Gabelli:2006eg,Feve:2007jx}  connected in series\cite{{Splettstoesser:2008gc},{Moskalets:2013dl},{Moskalets:2014ea}} or combined with an interferometer\cite{{Juergens:2011gu}}, or can be built from a single helical quantum capacitor\cite{{Inhofer:2013gu},{Hofer:2013cj}}. 
A quantum point contact with a time-dependent transmission can serve as a source of individual electron-hole pairs on-demand. \cite{Sherkunov:2009jm,Dasenbrook:2015bm} 
The most versatile multi-electron source is supposed to be the source of levitons proposed in Refs.~\onlinecite{Levitov:1996,Ivanov:1997,Keeling:2006} and experimentally realized in Ref.~\onlinecite{Dubois:2013dv}. 
This source uses a Lorentzian voltage pulse with quantized flux to generate excitations with an arbitrary integer charge for periodic \cite{Dubois:2013,Grenier:2013,Rech:2017,Vanucci:2017,Misiorny:2017,Ronetti:2017}  or random \cite{Glattli:2017vp} injection. 

\begin{figure}[t]
\includegraphics[width=80mm, angle=0]{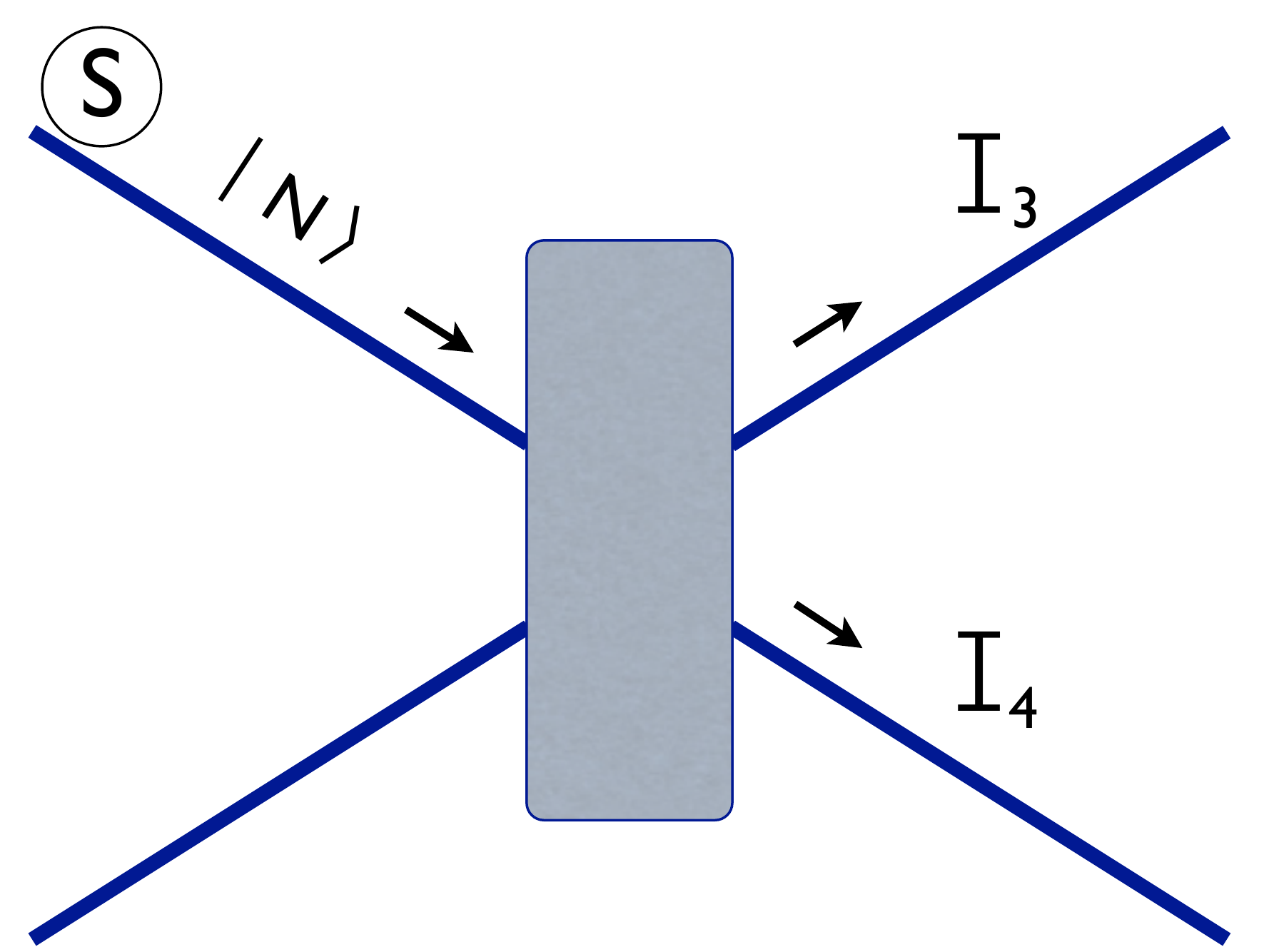}
\caption{A multi-particle electronic  state $\ket{N}$ emitted by the source {\bf S} and incoming to the wave splitter (a grey rectangle) gives rise the two outgoing currents, $I_3$ and $I_4$. The correlation function of these currents is used to gain information on the incoming quantum state. 
}
\label{fig1}
\end{figure}

A distinctive feature of the multi-particle system in comparison with a single particle is a possible presence of quantum-exchange correlations between  its constituents. 
These correlations can be detected with the help of the cross-correlation shot noise\cite{Buttiker:1990tn,Samuelsson:2004uv} measured in the geometry of the electron collider\cite{Buttiker:1992vr,Liu:1998wr}, see Fig.~\ref{fig1}. 
Notice the close analogy with the  famous in optics Hanbury Brown and Twiss effect\cite{HanburyBrown:1956bi}, which was also observed with electrons in mesoscopic conductors\cite{Henny:1999tb,Oliver:1999ws,Oberholzer:2000wx,Neder:2007jl}.  

In the present work I am interested in multi-electron systems created on the surface of the Fermi sea. 
The relevant source, for example, is based on quantum capacitors or is a source of levitons. 
The aim of the present work is to emphasize that the correlations between the particles composing an $N$-electron  system depend on the temperature of the underlying Fermi sea. 
To be definite, I analyze in detail a multi-electron state generated by the source of levitons. 

At zero temperature, the Pauli exclusion principle forces the electrons to be created in mutually orthogonal states with no quantum-exchange correlations to be present. \cite{Grenier:2013}
However, as the temperature increases, the state of electrons becomes mixed, and the correlations between them develop. 
These correlations enhance the shot noise compared to the one expected for individual electrons. 
Since the latter contribution scales as $N$ and the former contribution scales as $N^{2}$, the correlation shot noise dominates for large $N$. 
Moreover, in the limit of $N \to \infty$, the correlation noise completely destroys   the suppression of shot noise with temperature, the phenomenon, which was observed for single-particle\cite{Bocquillon:2012if,Glattli:2016tr} and two-particle\cite{Glattli:2016tr} injection. 

As it is well known, the correlations are not an intrinsic property of a multi-particle quantum state, but additionally they depend on the measurable of interest. 
This is why, in addition to charge noise, I analyze heat noise\cite{Krive:2001ci,Kindermann:2004im,Averin:2010kp,Sergi:2011eo,Zhan:2011ea,Sanchez:2013jx,{Crepieux:2014th}}, whose properties in the case a multi-electron system were addressed in Ref.~\onlinecite{Battista:2014tj}. 

Heat noise demonstrates properties that are both different and similar from those of charge noise. 
In particular, in contrast to charge noise, the heat noise of $N$-electron  system shows correlations already at zero temperature. 
On the other hand, at high temperatures, for both charge noise and heat noise, the correlation contribution to noise dominates for large $N$. 

Furthermore, for both charge and heat noise, the high-temperature asymptotics for multi-electron and  single-electron systems are the same. 
The latter fact allows us to consider a multi-electron system at high temperatures as an effective single particle.   
It looks like at high temperatures the exchange correlations fuse up electrons into one effective particle of total charge and energy.

The paper is organized as follows. 
In Sec.~\ref{sec2} the shot noise of a general multi-electron system at zero and nonzero temperatures is discussed. 
Sections \ref{sec3} and \ref{sec4} are devoted to a particular multi-electron  system, an $N$-electron leviton. 
Its charge and heat noise are considered in detail. 
The conclusion is given in Sec.~\ref{sec5}. 
Some technical details are given in Appendix~\ref{app1}.

\section{Shot noise in terms of the electronic correlation function}
\label{sec2}

The problem in question is the following. 
The electron source emits periodically a multi-particle state, which is divided  into two outgoing states by the wave splitter. 
In turn, these states give rise charge currents $I_{3}$ and $I_{4}$, see Fig.~\ref{fig1}. 
To access the possible correlations between electrons of a multi-particle state I  calculate the cross-correlation function of these currents integrated over the time difference $ \tau$ and averaged over the emission period $ {\cal T} $,\cite{{Pedersen:1998uc},{Moskalets:2004ct}}

\begin{eqnarray}
{\cal P}_{ 34} &=& 
\frac{1 }{2 } 
\int _{0}^{ {\cal T} } \frac{dt }{ {\cal T} } \int _{- \infty}^{ \infty } d \tau 
\nonumber \\
\label{01} \\
&&
\left\langle \Delta \hat I_{ 3}(t) \Delta \hat I_{ 4}(t + \tau)  + \Delta \hat I_{ 4}(t + \tau)  \Delta \hat I_{ 3}(t)  \right\rangle ,
\nonumber 
\end{eqnarray}
\ \\ \noindent
where $\Delta \hat I_{ \alpha}(t) =  \hat I_{ \alpha}(t) - \left\langle  \hat I_{ \alpha}(t) \right\rangle$, $ \alpha = 3,4$, is an operator of current fluctuations. 
The angle brackets  $\left\langle \dots  \right\rangle $ denote a quantum-statistical average over the equilibrium state of an incoming single-mode channel not affected by the electron source.  
Such an equilibrium state is the Fermi sea with a temperature $ \theta$ and a chemical potential $ \mu$. 
It is assumed that the other incoming channel of the wave splitter is in the same  equilibrium state, with the same temperature $ \theta$ and the same chemical potential $ \mu$. 

Fluctuations of a current are commonly called noise.
Fluctuations of a current caused by partitioning of electrons on a wave splitter  are commonly called the shot noise. \cite{Li:1990es,Liefrink:1991es,Reznikov:1995us,Blanter:2000wi}
For brevity, I call ${\cal P}_{ 34}$ the cross correlation shot noise or simply the shot noise.

To characterize the quantum state emitted by the source  I use the excess first-order electronic correlation function, \cite{Grenier:2011js,Grenier:2011dv,Haack:2013ch} 

\begin{eqnarray}
{ G}^{(1)}_{ }\left(1;2  \right) =
\langle \hat\Psi^{\dag}_{ }(1)  \hat\Psi_{ }(2) \rangle_{on} -  \langle \hat\Psi^{\dag}_{ }(1)  \hat\Psi_{ }(2) \rangle_{off},  
\label{02}
\end{eqnarray}
\ \\ \noindent
where $\hat\Psi_{ \alpha} (j) \equiv \hat\Psi_{ \alpha}\left(x_{j}t_{j} \right)$ is a single-particle electron field operator in second quantization evaluated at point $x_{j}$ and time  $t_{j}$,  $j=1,2$, after the source. 
The quantum statistical average $\langle \dots \rangle$ has the same meaning as in Eq.~(\ref{01}) and it is evaluated with the source switched on (the subscript ``on'') or off (the subscript ``off'').

Since here I am interested in quantum exchange correlations between electrons of a given multi-particle state, I suppose that the states emitted during different periods do not overlap with each other. 
This assumption allows us to treat formally the integral over period in Eq.~(\ref{01}) as an integral in infinite limits, $\int _{0}^{ {\cal T} } dt \to \int _{- \infty}^{ \infty } dt$. 
For simplicity, I omit the infinite limits of time integration below. 

For the collider set-up, the quantities ${\cal P}_{ 34}$, Eq.~(\ref{01}), evaluated at the output, and ${ G}^{(1)}_{ }$, Eq.~(\ref{02}), evaluated at the input, are related to each other in a simple manner, \cite{Moskalets:2015ub,Moskalets:2017dy} 

\begin{eqnarray}
\frac{  {\cal P}^{}_{34} }{  {\cal P}_{0} } &=& -  \iint\limits dt_{1} dt_{2}\left | v_{ \mu}^{}   G ^{(1)}_{}\left(t_{1};t_{2}\right)  \right |^{2} .
\label{03} 
\end{eqnarray}
\ \\ \noindent
Here the constant ${\cal P}_{0} = e^{2}T\left(  1-T \right)/ {\cal T}$ is interpreted as the shot noise caused by scattering off the wave splitter a single electron per period ${\cal T}$ at zero temperature.  
The factor $v_{ \mu}^{}$, the velocity of an electron with Fermi energy, comes from the normalization of a fermionic correlation function.
$G ^{(1)}_{}$ is  the excess first-order electronic correlation function evaluated just after the source. 
This is why I drop the coordinates $x_{1}= x_{2}$ and keep only the  times $t_{1}$ and $t_{2}$. 

The equation above was obtained within the Floquet scattering matrix approach under the following approximations. 
First, the wide band approximation was used. 
This approximation implies that the Fermi energy $ \mu$ is the dominant energy scale.
And second, the transmission probability $T$ of the wave splitter is supposed to be independent of energy (within the relevant energy interval, which is of the order of the energy of injected excitations).

\subsection{Zero temperature}

At zero temperature, the particles emitted by the source are in a pure quantum state. 
In this case, the correlation function of a multi-particle state with $N$ electrons is represented as follows, \cite{Grenier:2013,{Moskalets:2015vr}}

\begin{eqnarray}
v_{ \mu}^{}  G^{(1)}_{  }( t_{1};t_{2} ) &=&  
\sum\limits_{j=1}^{N}  \Psi^{*}_{j}(t_{1}) \Psi_{j}(t_{2}) ,
\label{04}  
\end{eqnarray}
\ \\ \noindent
where the wave functions are chosen to be orthonormal, 

\begin{eqnarray}
\int _{}^{ } dt \Psi^{*}_{j}(t_{}) \Psi_{j ^{\prime}}(t_{}) = \delta_{j j ^{\prime}}.
\label{05}
\end{eqnarray}
\noindent \\
Here $ \delta_{j j ^{\prime}}$ is the Kronecker delta. 

After substituting Eq.~(\ref{04}) into Eq.~(\ref{03}) we represent the shot noise as a  sum,
${\cal P}^{}_{34} = {\cal P}^{ind}_{34} + {\cal P}^{corr}_{34}$, 
where the first term, ${\cal P}^{ind}_{34}$, is the shot noise caused by scattering of individual electrons, 

\begin{eqnarray}
\frac{  {\cal P}^{ind}_{34} }{  {\cal P}_{0} } &=& 
-  \sum\limits_{j =1}^{N}
\left |  \int _{}^{ } dt \left | \Psi_{j }^{}(t) \right |^{2} \right |^{2} . 
\label{07} 
\end{eqnarray}
\noindent \\
The second term, $ {\cal P}^{corr}_{34}$, is an additional contribution to noise due to possible correlations between electrons, 

\begin{eqnarray}
\frac{  {\cal P}^{corr}_{34} }{  {\cal P}_{0} } &=& 
- \sum\limits_{j =1}^{N}
\sum\limits_{j ^{\prime} \ne j =1}^{N}
\left |  \int _{}^{ } dt \Psi_{j }^{*}(t) \Psi_{j ^{\prime} }^{}(t) \right |^{2} .
\label{08} 
\end{eqnarray}
\noindent \\
These correlations are quantified by the overlap integral between different  wave functions and they are detected as a deviation of the noise from the expected sum of individual contributions. 
In other words, the presence of correlations between electrons is naturally detected as non-additivity of the shot noise.  

Actually, at zero temperature, there are no any correlations. 
Using Eq.~(\ref{05}) we find, ${\cal P}^{ind}_{34} = - N {\cal P}_{0}$ and ${\cal P}^{corr}_{34}=0$. 
However, at nonzero temperatures this is not so.

\subsection{Nonzero temperatures}


Nonzero temperatures modify the state of the particles emitted by the source. 
Within the Floquet scattering approach, the excess correlation function of particles emitted at nonzero temperatures can be expressed in terms of the scattering amplitude of the source, $S_{in}\left( t,E \right)$.\cite{Moskalets:2015ub} 

Of special interest is the source for which the dependence of $S_{in}$ on two variables, enenrgy $E$ and time $t$, can be expressed in terms of one variable $E - ct$.  In such a case, one can avoid explicit use of the characteristics of the source  and express the excess correlation function at a nonzero temperature directly in terms of an  excess correlation function at zero temperature.\cite{Moskalets:2017vk}   

The source of levitons, realized in Ref.~\onlinecite{Dubois:2013dv}, is an example of such a source.  
Another example is the model of Ref.~\onlinecite{Keeling:2008ft}. 
In this model, electron emission occurs when a single quantum level, whose  energy varies linearly with time, crosses the Fermi energy. 
In particular, this model describes a source based on a quantum capacitor  when it is driven by a time dependent sine wave excitation applied to a metallic top gate, as in Ref.~\onlinecite{Bocquillon:2013fp}. 

In such a case,  
if the state emitted by the source at zero temperature is described by Eq.~(\ref{04}), then the state emitted at a nonzero temperature $ \theta$ is described by the following correlation function, 

\begin{eqnarray}
v_{ \mu} G^{(1)}_{ \theta  }( t_{1};t_{2} ^{}) =  
\sum\limits_{j=1}^{N}  
\int\limits_{- \infty}^{ \infty} d \epsilon^{} 
\left( - \frac{   \partial f}{  \partial \epsilon^{}  }  \right)
\Psi^{*}_{j, \epsilon}(t_{1}) \Psi_{j, \epsilon}(t_{2}^{}) ,
\label{09} 
\end{eqnarray}
\ \\ \noindent
where $f\left(  \epsilon \right) = \left(  1 + e^{\frac{  \mu + \epsilon }{ k_{B} \theta }} \right)$ is the Fermi distribution function, $k_{B}$ is the Boltzmann constant. 

The correlation function above tells us that the source now emits particles that are not in a pure state. 
The $j$th electron is now in a mixed quantum state with probability density $p_{ \theta}\left(  \epsilon \right) = -  \partial f/  \partial \epsilon $ and with component wave functions $ \Psi^{}_{j, \epsilon}$.  

Substituting Eq.~(\ref{09}) into Eq.~(\ref{03}) we find, ${\cal P}^{}_{34} = {\cal P}^{ind}_{34} + {\cal P}^{corr}_{34}$, where now the individual and correlation noise are defined as follows,

\begin{eqnarray}
\frac{  {\cal P}^{ind}_{34} }{  {\cal P}_{0} } \!\!  &=&  - 
\int\limits _{ - \infty }^{ \infty } d \epsilon  p_{\theta}( \epsilon) 
\int\limits _{ - \infty }^{ \infty } d \epsilon^{\prime} p_{\theta}( \epsilon^{\prime})  
\sum\limits_{j =1}^{N}
J_{j,j}\left(  \epsilon, \epsilon ^{\prime} \right),
\label{10}  
\end{eqnarray}
\ \\ \noindent 
and 

\begin{eqnarray}
\frac{  {\cal P}^{corr}_{34} }{  {\cal P}_{0} }  &=&  - 
\int\limits _{ - \infty }^{ \infty } d \epsilon  p_{\theta}( \epsilon) 
\int\limits _{ - \infty }^{ \infty } d \epsilon^{\prime} p_{\theta}( \epsilon^{\prime})  
\sum\limits_{j =1}^{N}
\sum\limits_{j ^{\prime} \ne j =1}^{N}
J_{j,j ^{\prime}}\left(  \epsilon, \epsilon ^{\prime} \right),
\nonumber \\
\label{11}
\end{eqnarray}
\ \\ \noindent
with the overlap integral squared being

\begin{eqnarray}
J_{j,j ^{\prime}}\left(  \epsilon, \epsilon ^{\prime} \right) &=&
\left |  \int _{}^{ } dt \Psi_{j, \epsilon}^{*}(t) \Psi_{j ^{\prime}, \epsilon ^{\prime}}^{}(t) \right |^{2} .
\label{11-1}
\end{eqnarray}
\ \\ \noindent 
Importantly, the component wave functions at different energies, $ \epsilon\ne \epsilon ^{\prime}$, are not orthogonal any longer. 
As a result, the correlation noise arises, ${\cal P}^{corr}_{34} \ne 0$. 

To be specific, on further on I focus on the source of levitons.

\section{Shot noise of an $N$-electron  leviton}
\label{sec3}

A voltage pulse of the Lorentzian shape, $eV(t) = N 2 \hbar \Gamma _{\tau} \left( t^{2} + \Gamma _{\tau}^{2}  \right)^{-1}$, applied to an Ohmic contact of a single-channel ballistic conductor, creates an $N$-electron  leviton of a time extent $2 \Gamma _{\tau}$ centered at $t=0$ (and propagating through the conductor with velocity $v_{ \mu}$). 
It is described by the correlation function of the form given in Eq.~(\ref{09}) with the  following set of $N$ single-particle wave functions, \cite{Grenier:2013,Moskalets:2015vr,Glattli:2016tr,{glattli2016method},Moskalets:2017vk} 

\begin{eqnarray}
\Psi_{j, \epsilon}(t) &=&  e^{-i t \frac{ \mu + \epsilon }{ \hbar} } \psi_{j}\left(  t \right) , \quad j = 1, \dots, N,
\nonumber \\
\label{12} \\
\psi_{j}\left(  t \right) &=&
\sqrt{\frac{\Gamma _{\tau}}{  \pi  } }
\frac{ 1 }{  t  - i \Gamma _{\tau} }
\left(  \frac{ t + i \Gamma _{\tau}  }{  t  - i \Gamma _{\tau}  } \right) ^{j-1} .
\nonumber 
\end{eqnarray}
\ \\ \noindent
I use the equation above in Eqs.~(\ref{10}) and (\ref{11}) and calculate  noise. 

The individual contribution is the same for all particles at nonzero  temperatures, as it was at zero temperature, 

\begin{eqnarray}
{\cal P}^{ind}_{34}  &=& -N {\cal P}^{1}_{34}, 
\nonumber \\
\label{13} \\
\frac{  {\cal P}^{1}_{34} }{  {\cal P}_{0} } &=&  
\int\limits _{ - \infty }^{ \infty } d \epsilon  p_{\theta}( \epsilon) 
\int\limits _{ - \infty }^{ \infty } d \epsilon^{\prime} p_{\theta}( \epsilon^{\prime})   e^{- \frac{ \left | \epsilon - \epsilon^{\prime} \right | }{  {\cal E}_{L}  }},
\nonumber 
\end{eqnarray}
\noindent \\
where ${\cal E}_{L} = \hbar / \left(  2 \Gamma _{\tau}  \right) $ is the energy of a $1$-electron leviton \cite{Keeling:2006}. 

Importantly, the exponential factor in Eq.~(\ref{13}) is the same for all the electrons constituting an $N$-electron  leviton. 
This is so, since the overlap integral $J_{j,j}$ entering Eq.~(\ref{10}) is independent of $j$. 
It depends only on the density profile, which is dictated by the shape of the voltage pulse used to create a leviton. 
  
\begin{figure}[t]
\includegraphics[width=80mm, angle=0]{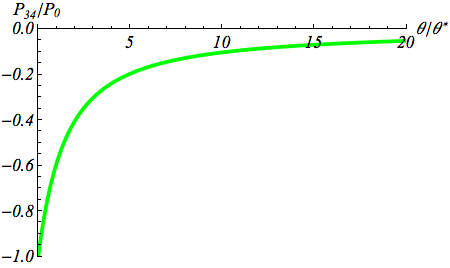}
\caption{
The temperature dependence of a single-particle shot noise,
${\cal P}^{}_{34} = -{\cal P}^{1}_{34}$, Eq.~(\ref{13}). 
The parameter $ k_B \theta^{*} = {\cal E}_{L}/ \pi$. 
}
\label{fig2}
\end{figure}

The temperature dependence of ${\cal P}^{1}_{34}$ is shown in Fig.~\ref{fig2}. 
So, the individual contribution to noise decreases with temperature.\cite{Moskalets:2017dy} 
The corresponding characteristic temperature depends on the time extent of the voltage pulse, $ k_B \theta^{*} = {\cal E}_{L}/ \pi =  \hbar/ \left( 2 \pi  \Gamma _{\tau} \right)$.

Worth mentioning that the density profile does not characterize fully a leviton. 
For instance, the energy of an $N$-electron  leviton is $N^{2} {\cal E}_{L}$, not $ {\cal E}_{L}$. \cite{{Moskalets:2009dk},Battista:2014tj}
However, the decay of the individual contributions to shot noise with temperature ${\cal P}^{ind}$, Eq.~(\ref{13}), does not reveal this energy.
This fact suggests that presumably the correlation noise becomes  dominant with increasing temperature. 
And indeed it is. 

\begin{figure}[t]
\includegraphics[width=80mm, angle=0]{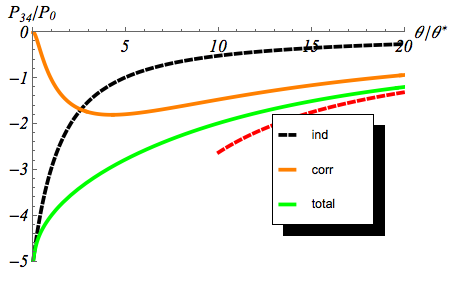}
\caption{
The temperature dependence of the shot noise of a $5$-electron leviton.
The individual contribution is given in Eq.~(\ref{13}) with $N=5$. 
The correlation contribution is defined in Eqs.~(\ref{11}), (\ref{11-1}), and (\ref{12}) with $N=5$. 
The dashed red line represents the high-temperature asymptotics of the total noise, see Eq.~(\ref{14}) with $N=5$.  
}
\label{fig3}
\end{figure}

In Fig.~\ref{fig3} the noise of a leviton with $N=5$ is shown. 
It is clear, that  rather rapidly with increasing temperature the correlation contribution to noise, ${\cal P}^{corr}$, Eq.~(\ref{11}), predominates the individual contribution.  

When considering the high-temperature noise asymptotics, the dominant role of correlations becomes even more pronounced.

\subsection{High temperatures}

At $k_{B} \theta \gg \max\left(  j,j ^{\prime} \right) {\cal E}_{L}$, the overlap integral can be approximated as follows, $J_{j,j ^{\prime}}\left(  \epsilon, \epsilon ^{\prime} \right) \approx 2 {\cal E}_{L} \delta\left(  \epsilon - \epsilon ^{\prime} \right)$. 
Note that in this limit the overlap integral does not depend on $j$, it is the same for $j=j ^{\prime}$ and for $j \ne j ^{\prime}$. 
In other words, in the high-temperature limit the mixed state of any electron (of the system in question) becomes effectively the same.

With this universal overlap integral, the shot noise, Eqs.~(\ref{10}) and (\ref{11}), is calculated rather trivially,

\begin{eqnarray}
\frac{ 1 }{ N }\frac{  {\cal P}^{ind}_{34} }{  {\cal P}_{0} } &\approx& -  \frac{ 1 }{ 3 } \frac{ {\cal E}_{L} }{ k_{B} \theta  }, \quad \quad k_{B} \theta \gg {\cal E}_{L}, 
\nonumber \\
\frac{ 1 }{ N } \frac{  {\cal P}^{corr}_{34} }{  {\cal P}_{0} } &\approx& -  \frac{ 1 }{ 3 } \frac{\left(  N - 1 \right) {\cal E}_{L} }{ k_{B} \theta  }, \quad \quad k_{B} \theta \gg N{\cal E}_{L}, 
\label{14} \\
\frac{ 1 }{ N } \frac{  {\cal P}^{}_{34} }{  {\cal P}_{0} } &\approx& - \frac{ 1 }{ 3 } \frac{ N {\cal E}_{L} }{ k_{B} \theta  }, \quad \quad k_{B} \theta \gg N{\cal E}_{L}. 
\nonumber 
\end{eqnarray}
\noindent \\
Here I show the noise per particle, ${\cal P}^{}_{34}/N$, instead of the total noise, ${\cal P}^{}_{34}$, by two reasons. 
First, to emphasize that the noise is not additive. 
Second, to show that the high-temperature noise is reduced compared to zero-temperature noise.
At zero temperature the shot noise per particle is simply $- {\cal P}_{0}$.  

\begin{figure}[t]
\includegraphics[width=80mm, angle=0]{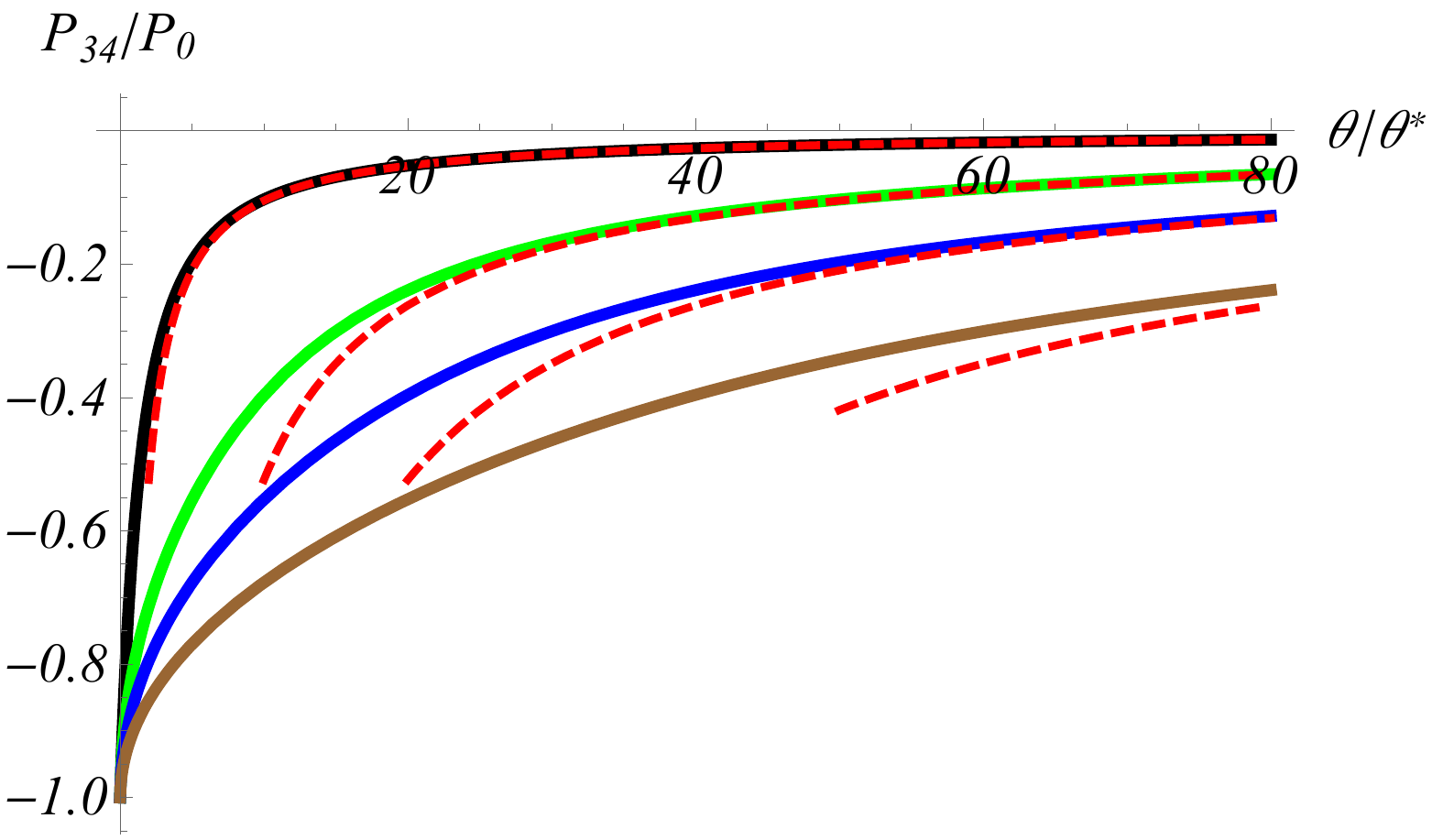}
\caption{
The temperature dependence of the shot noise per particle of an $N$-electron  leviton is shown for different $N=1,5,10,20$ (from top to bottom).
The dashed red lines represent the corresponding high-temperature asymptotics, see Eq.~(\ref{14}).  
}
\label{fig3-1}
\end{figure}

In the case of $N=5$, the corresponding asymptotics is shown in Fig.~\ref{fig3} as a dashed red line.
In Fig.~\ref{fig3-1} I show the temperature dependence of the shot noise (per particle) of a leviton with different numbers of electrons, $N$. 
This figure illustrates that with increasing $N$ the crossover to high-temperature behaviour occurs at higher temperatures, in agreement with Eq.~(\ref{14}). 

Equations (\ref{14}) show that in the case of a large number of particles,  the high-temperature correlation noise dominates, $\lim_{N \to \infty} {\cal P}^{corr}_{34}/{\cal P}^{ind}_{34} \sim N \to \infty$. 
As a result, the noise becomes quadratic in the number of particles. 
Since the noise in itself is quadratic in charge, ${\cal P}_{0} \sim e^{2}$, one can conclude that the high-temperature noise tells us about the total charge, $\left | {\cal P}^{}_{34} \right |_{ \theta \gg N {\cal E}_{L}} \sim \left(  eN \right)^{2}$. 
In contrast, the zero-temperature noise tells us about the total number of particles, $\left | {\cal P}^{}_{34} \right |_{ \theta \ll {\cal E}_{L}} \sim N$\cite{Bocquillon:2012if,Dubois:2013dv} or about the charge of individual quasi-particles\cite{Saminadayar:1997dw,dePicciotto:1997dk}. 

\begin{figure}[t]
\includegraphics[width=80mm, angle=0]{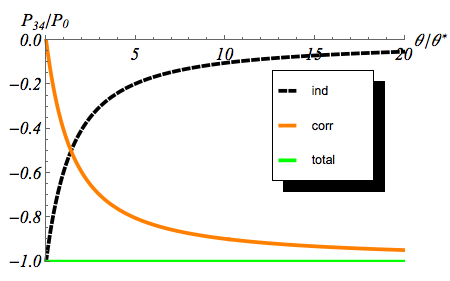}
\caption{
The temperature dependence of the shot noise per particle in the macroscopic limit, $N\to \infty$. 
The individual contribution is given in Eq.~(\ref{13}): $\lim_{N \to \infty} {\cal P}^{ind}_{34}/N = -{\cal P}^{1}_{34}$. 
The correlation contribution is given in Eq.~(\ref{15}). 
The parameter $ k_B \theta^{*} = {\cal E}_{L}/ \pi$. 
}
\label{fig4}
\end{figure}

Another important thing that we can learn from Fig.~\ref{fig3} is that there are two energy scales that determine the temperature dependence of the noise. 
One of them, $\left(  2N - 1 \right) {\cal E}_{L}$, is the Fermi energy of an $N$-electron  leviton, that is, the energy of a particle with an envelope wave function $ \psi_{N}$ (with the largest number)  from Eq.~(\ref{12}). 
The Fermi energy determines the crossover to the high-temperature regime, see Eq.~(\ref{14}) and dashed red lines in Figs.~\ref{fig3} and \ref{fig3-1}.  
The other energy, $ 2{\cal E}_{L}$, is an analogue of the level spacing for this multi-particle system. 
The level spacing determines the low-temperature crossover from individual to correlation noise, see the intersection in Figs.~\ref{fig3} and \ref{fig4}.

\subsection{The macroscopic limit}

An interesting effect takes place at intermediate temperatures, $ {\cal E}_{L} \ll k_{B} \theta \ll N {\cal E}_{L}$ with an increase in the number of particles $N$. 
On one hand, the correlation noise dominates there and it is expected to grow as $N^{2}$. 
On the other hand, the noise suppression is not strong at these temperatures and one might expect that the noise at a nonzero temperature can overcome the zero-temperature noise, which grows only as $N$.

Actually this is a wrong expectation. 
The shot noise at zero temperature sets the upper bound to the noise caused by partitioning of an electron, regardless of whether it is single or belongs to a multi-particle system, like an $N$-electron  leviton. 
And this upper bound is perfectly reached in the macroscopic limit, $N \to \infty$. 
Let us show this. 
In order to deal with finite quantities we calculate the noise per particle. 

The individual contribution to noise ${\cal P}^{ind}_{34}$, Eq.~(\ref{13}), holds for any $N$. 
The correlation contribution in the macroscopic limit is, (see Appendix~\ref{app1})  

\begin{eqnarray}
\lim_{N \to \infty} \frac{ 1 }{ N } \frac{  {\cal P}^{corr}_{34} }{  {\cal P}_{0} }  &=& -1 + \frac{  {\cal P}^{1}_{34} }{  {\cal P}_{0} }.
\label{15}
\end{eqnarray}
\noindent \\
Therefore, in the macroscopic limit, $N\to \infty$, the noise per particle is independent of temperature, $\lim_{N \to \infty}{\cal P}^{}_{34}/N = \lim_{N \to \infty}\left(  {\cal P}^{ind}_{34} + {\cal P}^{corr}_{34} \right)/N = -  {\cal P}^{}_{0}$.

The total, individual, and correlation noise per particle in the macroscopic limit are shown in Fig.~\ref{fig4} as a function of temperature. 
Interestingly, the low-temperature crossover from individual to correlation noise is perfectly hidden, since the measured noise, ${\cal P}^{}_{34}$, does not change at all at this temperature.

\subsection{Experimental relevance}

In principle, both low- and high-temperature regimes are accessible in current day experiments. 
For instance, in Ref.~\onlinecite{Glattli:2016tr} the levitons with $ 2  \Gamma _{\tau} = 75 {\rm ps}$ were generated at various temperatures, $\theta= 40 {\rm mK}, 90 {\rm mK}$ and $138 {\rm mK}$. 
For such excitations $ {\cal E}_{L} \approx 320 {\rm mK}$ and the characteristic temperature is $ \theta^{*} \approx 100  {\rm mK}$. 
From Fig.~\ref{fig2} we see that at the lowest temperature, $\sim 0.4 \theta^{*}$, a single-electron leviton is in the low-temperature regime, where its properties are almost not affected by temperature. 
At an intermediate temperature, $\sim 0.9 \theta^{*}$, the shot noise is reduced by approximately $30\%$. 
Therefore, we conclude that such a temperature already has a significant effect on the state of a leviton. 
From Fig.~\ref{fig3-1} (the topmost line) we see that at the highest temperature used in Ref.~\onlinecite{Glattli:2016tr}, $\sim 1.4 \theta^{*}$, the shot noise of a single-electron leviton is approaching its high-temperature asymptotics (see a corresponding dashed red line). 

The state of an $N$-electron  leviton is also affected by the temperatures in this range. 
In particular, we see  from Figs.~\ref{fig3} that for $N=5$ the correlation noise exceeds a single-particle noise at $ \theta > 2.5 \theta^{*} \approx 250 {\rm mK}$. 
This fact tells us that at such high temperatures the states of the individual particles composing a leviton are not orthogonal to each other (at zero temperature they are orthogonal to each other).  
With increasing $N$ the temperature of the crossover from individual to correlation noise becomes even smaller, $\sim 1.5 \theta^{*} \approx 150 {\rm mK}$ for $N \to \infty$, see the intersection in Fig.~\ref{fig4}. 

To observe a single-particle-like behaviour of an $N$-electron  leviton, which manifests itself in universal high-temperature shot noise, Eq.~(\ref{14}), somewhat higher temperatures should be used. 
In particular, for $N=5, 10$, and $20$, the temperature should exceed  $\sim 9 {\cal E}_{L}( \approx 3 {\rm K})$, $\sim 19 {\cal E}_{L}(\approx 6 {\rm K})$, and  $\sim 39 {\cal E}_{L}(\approx 12 {\rm K})$, respectively, see Fig.~\ref{fig3-1} and remember that $  {\cal E}_{L} = \pi k_{B} \theta^{*}$.

\section{Heat noise of an $N$-electron  leviton}
\label{sec4}

In this section I consider the heat noise caused by the scattering of an $N$-electron  leviton off a wave splitter, as shown in Fig.~\ref{fig1}.   
The aim of this section is to show the differences and similarities between charge and heat noise. 

In particular, in contrast to a charge noise, the heat noise contains a correlation contribution already at zero temperature. 
On the other hand, at high temperatures both charge and heat noise can be interpreted as if they are caused by one (indivisible) particle with total  charge and energy. 

\subsection{Heat noise in terms of the electronic correlation function}

The cross-correlation function of heat currents, ${\cal P}_{ 34}^{Q} $ is defined by Eq.~(\ref{01}), where the operator of a charge current $\hat I_{ \alpha}$ is replaced by the operator of a heat current $\hat I_{ \alpha}^{Q}$.\cite{{Battista:2013ew},Ludovico:2014de} 
By analogy with a charge noise, Eq.~(\ref{03}), the heat noise can be expressed in terms of the excess first-order electronic correlation function ${ G}^{(1)}_{ }$ of the excitations generated by the source, as follows, \cite{{Moskalets:2016va}}

\begin{eqnarray}
\frac{  {\cal P}^{Q}_{34} }{  {\cal P}_{0}/e^{2} } &=&  - 
\iint _{}^{ }  dt_{1} d t_{2} 
\left( - i \hbar  \frac{\partial  }{\partial t_{1} } - \mu \right)v_{ \mu}G^{(1)}_{ }( t_{1};t_{2})  
\nonumber \\
\label{16} \\
&& \times
\left( - i \hbar  \frac{\partial  }{\partial t_{2} } - \mu \right)v_{ \mu}G^{(1)}_{ }( t_{2};t_{1})  
.
\nonumber 
\end{eqnarray}

In the case of an $N$-particle state, Eqs.~(\ref{04}), (\ref{09}), heat noise can be conveniently represented as the sum of individual and correlation contributions, ${\cal P}^{Q}_{34} = {\cal P}^{Q,ind}_{34} + {\cal P}^{Q,corr}_{34}$, which are given by Eqs.~(\ref{10}) and (\ref{11}), respectively, but with ${\cal P}_{0}$ being replaced by ${\cal P}_{0}^{}/e^{2}$ and the overlap integral $J_{j,j ^{\prime}}\left(  \epsilon, \epsilon ^{\prime} \right)$ being replaced by the following overlap integral, 

\begin{eqnarray}
J_{j, j ^{\prime}}^{Q}\left(  \epsilon, \epsilon ^{\prime} \right) &=&  
\left |  \int _{}^{ } dt \Psi_{j ^{\prime} , \epsilon ^{\prime}}^{}(t) \left( - i \hbar  \frac{\partial  }{\partial t_{} } - \mu \right)  \Psi_{j, \epsilon}^{*}(t)   \right |^{2} . 
\nonumber \\
\label{17}
\end{eqnarray}
\ \\ \noindent
Importantly, the charge and heat noise are determined by different overlap integrals.\cite{Moskalets:2016va} 
This is the mathematical reason why  their properties are different.

\subsection{Zero temperature}

At zero temperature, what matters is the overlap integral evaluated at $ \epsilon = \epsilon ^{\prime} =0$. 
With the wave functions from Eq.~(\ref{12}), the overlap integral becomes,

\begin{eqnarray}
J_{j, j ^{\prime}}^{Q}\left( 0,0 \right) &=& {\cal E}_{L}^{2} \left\{ \left( 2j - 1 \right)^{2} \delta_{j, j ^{\prime}} + \left( j - 1 \right)^{2} \delta_{j, j ^{\prime}\pm 1}  \right\} .
\nonumber \\
\label{18}
\end{eqnarray}
\ \\ \noindent  
First of all, what we see from this equation is that different particles contribute differently to heat noise. 
As a result, even individual contribution to heat noise is not additive,

\begin{eqnarray}
\frac{ {\cal P}^{Q,ind}_{34}  }{  {\cal P}_{0}^{}/e^{2} }  = - {\cal E}_{L}^{2}  \sum\limits_{j=1}^{N} \left( 2j-1 \right)^{2} = - {\cal E}_{L}^{2} \frac{ 4N^{3} -  N }{ 3 } .
\label{19}
\end{eqnarray}
\ \\ \noindent
This is in contrast to charge noise, where the individual contributions of different particles are the same. 
As a result, the charge noise is additive.  

This difference is rather trivial. 
The charge noise is proportional to the charge (squared), which is the same for all the particles.  
While the heat noise is proportional to the energy (squared), which is different for different particles: The energy of a particle with wave function $ \Psi_{j, \epsilon =0}$, Eq.~(\ref{12}), is $ {\cal E}_{j} = {\cal E}_{L}\left( 2j-1 \right)$.

Another conclusion derived from Eq.~(\ref{18}) is that despite the orthogonality of the different wave functions, see Eq.~(\ref{05}), the correlation contribution to heat noise does not vanish even at zero temperature, 

\begin{eqnarray}
\frac{ {\cal P}^{Q,corr}_{34}  }{  {\cal P}_{0}^{}/e^{2} }  = - {\cal E}_{L}^{2} 2 \sum\limits_{j=2}^{N} \left( j-1 \right)^{2} = - {\cal E}_{L}^{2} \frac{ 2N^{3} - 3 N^{2}+N }{ 3 } .
\nonumber \\
\label{20}
\end{eqnarray}
This is in strike contrast with charge noise, which has no a correlation contribution at zero temperature, ${\cal P}^{corr}_{34} = 0$, see Eqs.~(\ref{08}) and (\ref{05}).   
Such a difference is due to the fact that the wave functions $ \Psi_{j, \epsilon}$, Eq.~(\ref{12}), describe single-particle states with a fixed charge, but with a fluctuating energy \cite{Battista:2013ew} (the energy $ {\cal E}_{j}$, mentioned above, is the mean energy). 
Therefore, in this case the orthogonality of the wave functions does not by itself prevent the correlations between the energies carried by the particles.

Curiously, despite the disagreement over the effect of correlations, there is a  certain consensus between the total charge and heat noise at zero temperature:  
The total heat noise (the total charge noise) is proportional to the product of the total energy, $N^{2} {\cal E}_{L}$,  (of the total charge, $Ne$), and the energy of the topmost electron, $(2N-1) {\cal E}_{L}$, (the charge of an electron, $e$).

Note that for single-particle states with a fixed energy, when the time dependence of a wave function is purely exponential, $ \Psi(t) \sim e^{\frac{i }{ \hbar } \epsilon t}$, the charge and heat overlap integrals are proportional to each other, $J^{Q}_{j,j ^{\prime}} \sim J^{}_{j, j ^{\prime}} $. 
As a result, the correlation charge and heat noise would be simultaneously    either zero or nonzero.  

Interestingly, in our case the charge and heat overlap integrals are proportional to each other in the high-temperature limit, where the correlation contribution to noise is dominant for both charge and heat noise. 
However these overlap integrals depend on energy differently. 
As a result, the charge noise decreases as $1/ \theta$, see Eq.~(\ref{14}), while, as we shall show below, the heat noise increases as $\theta$, as expected \cite{Jarzynski:1997uj,Crooks:1999ta,Bunin:2011kx,Averin:2011cc,Bochkov:2013bq,Moskalets:2014cr}.

\subsection{High temperatures}

At sufficiently high temperatures, namely at $k_{B} \theta \gg {\cal E}_{j}$, the dominant contribution to heat noise comes from energies $ \epsilon \sim k_{B} \theta$. 
Therefore, while evaluating the overlap integral $J_{j, j ^{\prime}}^{Q}$, Eq.~(\ref{17}), we differentiate only the phase factor of the wave function $ \Psi_{j, \epsilon}$, Eq.~(\ref{12}), and obtain $\left( - i \hbar  \frac{\partial  }{\partial t_{} } - \mu \right)  \Psi_{j, \epsilon}^{*}(t) \approx \epsilon  \Psi_{j, \epsilon}^{*}(t)$. 
Differentiation of the prefactor of the wave function ultimately would result in the energy of a corresponding electron, ${\cal E}_{j}$, which we neglect. 
Then at $k_{B} \theta \gg \max\left(  {\cal E}_{j}, {\cal E}_{j ^{\prime}}\right)$ the heat overlap integral is expressed in terms of the electric overlap integral $J^{Q}_{j,j ^{\prime}}\left(  \epsilon, \epsilon ^{\prime} \right) = \epsilon^{2} J^{}_{j,j ^{\prime}}\left(  \epsilon, \epsilon ^{\prime} \right) \approx   \epsilon^{2} 2 {\cal E}_{L}  \delta\left(  \epsilon - \epsilon ^{\prime} \right)$. 
Using this approximation we find at $k_{B} \theta \gg N {\cal E}_{L}$,

\begin{eqnarray}
\frac{  {\cal P}^{Q,ind}_{34} }{  {\cal P}_{0}/e^{2} } &\approx& 
- N\frac{ \pi^{2} - 6 }{ 9  }{\cal E}_{L} k_{B} \theta ,  
\nonumber \\
\frac{  {\cal P}^{Q,corr}_{34} }{  {\cal P}_{0}/e^{2} } &\approx& 
- N\left( N-1 \right)\frac{ \pi^{2} - 6 }{ 9  }{\cal E}_{L} k_{B} \theta ,  
\label{21} \\
 \frac{  {\cal P}^{Q}_{34} }{  {\cal P}_{0}/e^{2} } &\approx& 
 - \frac{ \pi^{2} - 6 }{ 9  }  N^{2} {\cal E}_{L} k_{B} \theta .   
\nonumber 
\end{eqnarray}
\noindent \\ 
The individual contribution is the same for all particles, so it is additive, that is,  it is proportional to the number of particles $N$. 
The correlation contribution of each pair of particles is the same, so it is proportional to the number of pairs, $N(N-1)$. 

Interestingly, the amount of noise associated with one extra particle is proportional to the energy of this particle. 
In other words, when we add the $N$th particle to the system of $N-1$ particle, the high-temperature heat noise increases by a value proportional to $(2N-1) {\cal E}_{L}$. 
The predominant contribution here is due to correlation noise.

Note that the total heat noise is of the form of the heat noise of a single particle, in which the energy of a single particle $ {\cal E}_{L}$ is replaced by the total energy of an $N$-electron  leviton, $N^{2} {\cal E}_{L}$.  
This observation goes in line with a similar observation for charge noise. 
This suggests that at high temperatures the system of $N$ electrons can be regarded as one particle of total charge and energy. 

\section{Conclusion}
\label{sec5}

The effect of temperature on the shot noise of a multi-electron ($N$-electron) leviton is investigated. 
The physics behind such an effect consists in the fact that the  temperature of the underlying Fermi sea affects the quantum state of a leviton: While at zero temperate its elementary constituents (i.e., electrons having an elementary charge $e$)  are  created in a pure state, at finite temperature they are created in a mixed state. 
The pure states are mutually orthogonal, while the mixed states are not. 
The specific effect of orthogonality depends on the particular quantity of interest.
As examples, the charge cross-correlation noise and heat cross-correlation noise were considered. 
 
The charge noise behaves with temperature as follows. 
At zero temperature, the leviton's elementary constituents all are in mutually orthogonal states. 
As a result, each of them contribute independently to charge noise. 
All individual contributions are the same, since any electron carry the same charge $e$ (the noise is proportional to the charge squared). 
Therefore, the total charge noise at zero temperature is proportional to the number of leviton's constituents $N$. 

At a nonzero temperature, this is no longer the case.
The various components of mixed state of different electrons are not orthogonal to each other. 
As a result, the correlation contribution to noise arises. 
This contribution scales as $N^{2}$, and for large $N$ it becomes dominant over the sum of individual contributions. 
So, at high temperatures the charge shot noise is proportional to $N^{2}$,  which is expected for the shot noise of an indivisible particle with charge  $Ne$.  
The charge shot noise depends also on temperature. 
Its decrease with temperature is due to the mixedness of the quantum state.

The crossover from low- to high-temperature behavior depends on $N$. 
For $N=1$ the crossover temperature $ \theta^{*}$  is proportional to the energy of a single-electron leviton, ${\cal E}_{L} = \hbar/(2 \Gamma _{\tau})$, where $ 2\Gamma _{\tau}$ is the duration of a leviton.  
At higher $N$, the correlation contribution increases the crossover temperature, since it enhances noise. 
Now the crossover temperature is of the order of $(2N-1) {\cal E}_{L}$, which is the energy of the topmost electron, the Fermi energy of this multi-particle system.  
Moreover, at $N \to \infty$ the correlation contribution fully compensates for the reduction of single-particle noise with temperature. 

The behavior of heat noise is different. 
Since the pure quantum state of leviton's constituents is a superposition of states with different energies, already at zero temperature there is a  correlation contribution to heat noise. 
For large $N$, it is as large as half of the contribution of individual constituents. 
At high temperatures, in contrast to charge noise, the heat noise grows with temperature. 
The prefactor is proportional to $N^{2}$, namely to the total energy of a leviton.   

Though the overall behavior of charge and heat noise is different, they  both seem to demonstrate that at high temperatures the elementary constituents of a multi-electron leviton become effectively fused together. 
As a result, we see an effective single indivisible particle with total charge $Ne$ and total energy $N^{2} {\cal E}_{L}$. 
Responsible for this are the classical temperature-induced correlations between the various components of the mixed state of the multi-electron leviton.

\acknowledgments
I appreciate the warm hospitality of the Aalto University, Finland, where this project was started.

\appendix
\section{Shot noise of an $N$-electron  leviton in the macroscopic limit, $N\to \infty$}
\label{app1}

First, let us calculate the correlation noise ${\cal P}^{corr}_{34}$, Eq.~(\ref{11}), with wave functions $ \Psi_{j, \epsilon}$ from Eq.~(\ref{12}) at finite $N$. 
And then take a limit $N \to \infty$. 

For this purpose it is convenient to integrate out the energies $ \epsilon$ and $ \epsilon ^{\prime}$ in Eq.~(\ref{11}), 

\begin{eqnarray}
\frac{  {\cal P}^{corr}_{34} }{  {\cal P}_{0} } &=& -  \iint\limits dt_{1} dt_{2} 
\eta ^{2}\left( \frac{  t_{1} - t_{2}  }{ \tau_{ \theta} }  \right) {\cal J}\left( t_{1}, t_{2}  \right) ,
\nonumber \\
\label{a01} \\
{\cal J}\left( t_{1}, t_{2} \right)  &=& 
\sum\limits_{j =1}^{N}
\sum\limits_{j ^{\prime} \ne j =1}^{N}
\psi^{*}_{j}(t_{1}) \psi_{j}(t_{2}) 
\psi^{}_{j ^{\prime}}(t_{1}) \psi^{*}_{j ^{\prime}}(t_{2}) ,
\nonumber 
\end{eqnarray}
\noindent \\
where $ \eta\left( x \right) = x/ \sinh(x)$ and $\tau_{ \theta} =  \hbar /\left(   \pi k_{B} \theta \right)$. 
The product of four wave functions from the equation (\ref{12}) depends on the difference $m = j - j ^{\prime}$ rather than on $j$ and $j ^{\prime}$ separately. 
This fact allows us to reduce the double sum to a single sum, 

\begin{eqnarray}
{\cal J}\left( t_{1}, t_{2} \right)  &=& N \frac{2  \Gamma _{\tau}^{2} }{ \pi^{2} }
\sum\limits_{m=1}^{N-1} \frac{\left( 1 - \frac{m }{ N} \right)  {\rm Re} \left [ q^{m}  \right] }{\left( t_{1}^{2} +  \Gamma _{\tau}^{2} \right)\left( t_{2}^{2} +  \Gamma _{\tau}^{2} \right) },
\nonumber \\
\label{a02} \\
q &=&
\frac{ \left(  t_{1} - i \Gamma _{\tau} \right) ^{}  }{ \left(  t_{1} + i \Gamma _{\tau} \right) ^{}  } 
\frac{ \left(  t_{2} + i \Gamma _{\tau} \right) ^{}  }{  \left(  t_{2} - i \Gamma _{\tau} \right) ^{}  }  .
\nonumber 
\end{eqnarray}
\ \\ \noindent
In the macroscopic limit, $N \to \infty$, one can ignore $m/N$ compared to $1$ (accounting for this term leads to corrections $\sim 1/N$). 
%
%

To evaluate the resulting sum, I proceed as follows. 
I introduce the time difference, $ \tau = t_{1} - t_{2}$, and note that $q\ne 1$ for $ \tau \ne 0$. 
In this case the sum in Eq.~(\ref{a02}) can be evaluated as $\lim_{N \to \infty} \sum_{m=1}^{N-1} {\rm Re} \left [ q^{m}  \right] = {\rm Re} \left [\frac{ q }{ 1 - q } \right]$. 
We can use this result in Eq.~(\ref{a02}) and evaluate the integral in Eq.~(\ref{a01}) for any $ \tau \ne 0$. 

In contrast, for $ \tau = 0$ we have $q=1$ and the result above diverges, indicating that formally we cannot use it. 
Despite the fact that this divergency is integrable, that is, Eq.~(\ref{a01}) remains finite, we cannot rule out the possibility of existence of an additional contribution arising  solely for $ \tau=0$. 
Such a contribution, if it exists, manifests itself as a singularity, $\sim \delta\left( \tau \right)$, in the sum over $m$ in equation for the overlap integral ${\cal J}\left( t_{1}, t_{2} \right)$, Eq.~(\ref{a02}). 
So, the overlap integral can be represented in the general form as follows,

\begin{eqnarray}
\lim_{N \to \infty} \frac{ 1 }{ N } {\cal J}\left( t_{1}, t_{2} \right)  &=& 
\frac{ \alpha \delta\left( \tau \right) }{ \left( t^{2} +  \Gamma _{\tau}^{2} \right)^{2} } 
\label{a03} \\
&&
-  \frac{  \Gamma _{\tau}^{} }{ \pi^{2}  \tau } {\rm Im} \left [ 
\frac{  1 }{\left( t_{1}^{} +  i\Gamma _{\tau}^{} \right)\left( t_{2}^{} -i  \Gamma _{\tau}^{} \right) }  \right],
\nonumber 
\end{eqnarray}
\ \\ \noindent
where $t = \left( t_{1} + t_{2} \right)/2$ is the mean time and  
the constant $ \alpha$ to be determined. 
If we would find that $ \alpha=0$, then it would mean, that there is no a singular contribution. 
However, as I show below, $ \alpha \ne 0$ in the present case. 

Now I substitute Eq.~(\ref{a03}) into Eq.~(\ref{a01}), integrate out the mean time $t = \left( t_{1} + t_{2} \right)/2$, and get the correlation noise per particle, 

\begin{eqnarray}
\lim_{N \to \infty} \frac{ 1 }{ N } \frac{  {\cal P}^{corr}_{34} }{  {\cal P}_{0} } &=& 
 -  \frac{\alpha \pi  }{ 2  \Gamma _{\tau}^{3} } +
\frac{2  \Gamma _{\tau} }{ \pi } 
\int\limits  d \tau  
\frac{\eta ^{2}\left( \frac{ \tau  }{ \tau_{ \theta} }  \right) }{ \tau^{2} + 4  \Gamma _{\tau}^{2} } .
\nonumber \\
\label{a04}
\end{eqnarray}
\ \\ \noindent
Note, the singular term in Eq.~(\ref{a03}) results in a contribution, which is independent of temperature. 
Therefore, to evaluate the constant $ \alpha$, I can use the correlation noise at zero temperature, which was already calculated. 
According to Eqs.~(\ref{08}) and (\ref{05}), at zero temperature the correlation noise vanishes for any $N$. 
At $ \theta=0$, when $ \eta\left( \tau/ \tau_{ \theta} \right) = 1$, the equation above is nullified if $ \alpha = 2 \Gamma _{\tau}^{3}/ \pi$. 

Easy to check that the second term on the right hand side of Eq.~(\ref{a04}) is exactly the same as ${\cal P}^{1}_{34} / {\cal P}_{0}$, given in Eq.~(\ref{13}). 
Therefore, Eq.~(\ref{a04}) is nothing but Eq.~(\ref{15}).

\end{document}